\documentclass[aps,prb,twocolumn,showpacs]{revtex4}

\usepackage{graphicx}
\usepackage{amsmath}
\usepackage{amssymb}

\begin{document}

\title{Extinction of quasiparticle scattering interference in
cuprate superconductors}

\author{Zhi Wang}
\affiliation{Department of Physics, Beijing Normal University,
Beijing 100875, China}

\author{Bin Liu}
\affiliation{Department of Physics, Beijing Normal University,
Beijing 100875, China}

\author{Shiping Feng\footnote{
Corresponding author. Tel.: +86 10 58806408; Fax: +86 10 58801764;
E-mail address: spfeng@bnu.edu.cn}}

\affiliation{Department of Physics, Beijing Normal University,
Beijing 100875, China}

\begin{abstract}
The quasiparticle scattering interference phenomenon characterized
by the peaks in the local density of states is studied within the
kinetic energy driven superconducting mechanism in the presence of a
single impurity. By calculation of the Fourier transformed ratio of
the local density of states at opposite energy, it is shown that the
quasiparticle scattering interference phenomenon can be described
qualitatively by a single impurity in the kinetic energy driven
homogeneous d-wave superconducting state. The amplitude of the peak
increases with increasing energy at the low energy, and reaches a
maximum at the intermediate energy, then diminishes to zero at the
high energy. The theory also predicts that with increasing doping,
the position of the peak along the nodal direction moves towards to
the center of the Brillouin zone, while the position of the peak
along the antinodal direction is shifted to large momentum region.
\end{abstract}


\maketitle

In the study of the physical properties of cuprate superconductors,
one of powerful techniques is scanning tunneling microscopy (STM)
\cite{fischer07}, since it is the only method to probe the
real-space inhomogeneous electronic structure of cuprate
superconductors. Application of STM to cuprate superconductors would
allow one to explore some important issues
\cite{fischer07,pan01,hoffman02,mcelroy03}, including (1) the
physical processes dominating quasiparticle scattering, (2) the
degree to which the quasiparticles are well defined and coherent,
and (3) the relation between the commensurate and incommensurate
magnetic signatures and the quasiparticle scattering processes. In
particular, STM has been used to infer the momentum-space
information of quasiparticles from the Fourier transform of the
position (${\bf r}$) and energy ($\omega$)-dependent local density
of states (LDOS), $\rho({\bf r},\omega)$, then both real-space and
momentum-space modulations for LDOS are explored simultaneously
\cite{hoffman02}. The typical feature observed by the Fourier
transformed LDOS is dominated by the peaks at well-defined
wavevectors ${\bf q}_{i}$ obeying the octet model
\cite{pan01,hoffman02,mcelroy03}, since the quasiparticle dispersion
in cuprate superconductors in the superconducting (SC) state has
closed constant-energy contours surrounding the d-wave nodes.
Theoretically, there is a general consensus that the peak emerges
due to the quasiparticle scattering interference (QSI)
\cite{wang03,zhang03,capriotti03,zhu04}. This QSI is manifested
itself as spatial modulation in the Fourier transformed LDOS
$\rho({\bf q},\omega)$ whose wavevectors change with energy. Several
attempts have been made to make this argument more precise. In
particular, some calculations based on the phenomenological d-wave
Bardeen-Cooper-Schrieffer (BCS) formalism have been performed by
considering the effect of impurity scattering
\cite{wang03,zhang03,capriotti03,barnea03,zhu04,misra04}, where it
has been shown that a single or few impurities in a homogeneous
d-wave SC state leads to a result that is in qualitative agreement
with the STM experimental data \cite{pan01,hoffman02,mcelroy03}.

Recently, the improvements in the resolution of STM experiments
\cite{kohsaka07,hanaguri07,kohsaka08} allowed one to resolve
additional features in QSI. Among these new achievements is the
measurement of the ratio of differential conductances at opposite
bias,
\begin{eqnarray}\label{ratio}
Z({\bf r},V)\equiv g({\bf r},+V)/g({\bf r},-V),
\end{eqnarray}
where $V$ is the bias voltage and $g({\bf r},V)$ the differential
conductance. The advantage of this procedure is that it cancels the
severe systematic errors in the measurement of $g({\bf r},+V)$ due
to tip elevation errors \cite{kohsaka07,hanaguri07,kohsaka08}, yet
retains all the quasiparticle information in the differential
conductance. In particular, an remarkable phenomenon observed from
these new STM experiments is that QSI always disappears at a scale
energy (high energy) that is indistinguishable from the energy at
which electronic homogeneity is lost. In other words, at energies
below the high energy, the STM response is similar to the dispersing
Bogoliubov quasiparticle, however, at energies above the high
energy, the response becomes highly spatially inhomogeneous. In
corresponding to the constant-energy contour of the high energy, an
extinction line exists \cite{kohsaka08} in momentum space [near the
diagonal line between ($\pm\pi,0$) and ($0,\pm\pi$)], beyond which
most of the dispersing peaks disappear, to be replaced by a reduced
set of roughly non-dispersive peaks. These new STM experimental
results lead naturally to speculation about the appearance of the
QSI extinction in cuprate superconductors in the SC state near the
boundary of the Brillouin zone (antinodal regime). Within the
phenomenological BCS formalism, the QSI phenomenon in a d-wave SC
state with coexisting short-range antiferromagnetic (AF) order has
been studied \cite{anderson09,atkinson05} recently by considering
the scattering arising from a single point-like impurity, and the
result shows that dispersing peaks are then extinguished at the high
energy \cite{anderson09}. To the best of our knowledge, the
dispersing peaks below the high energy in cuprate superconductors
and its extinction at the high energy has not been treated starting
from a microscopic SC theory, and no explicit calculations of the
doping dependence of the dispersing peaks has been made so far. In
this paper we show very clearly if the effect of a single point-like
impurity scattering potential is considered within the kinetic
energy driven SC mechanism \cite{feng03}, the calculated ratio of
LDOS at opposite energy in momentum-space can reproduce some main
features observed experimentally on cuprate superconductors
\cite{hoffman02,mcelroy03,kohsaka08}, including the appearance of
the dispersing peaks below the high energy and its extinction at the
high energy. With increasing doping concentration, the position of
the peak along the nodal direction moves towards to the center of
the Brillouin zone, while the position of the peak along the
antinodal direction is shifted to large momentum region.

In cuprate superconductors, the single common feature is the
presence of the two-dimensional CuO$_{2}$ plane, and then it is
believed that the unconventional physical properties of cuprate
superconductors is closely related to the doped CuO$_{2}$ planes
\cite{shen03}. It has been argued that the $t$-$J$ model on a square
lattice,
\begin{eqnarray}\label{t-j1}
H&=&-t\sum_{i\hat{\eta}\sigma}C^{\dagger}_{i\sigma}
C_{i+\hat{\eta}\sigma}+t'\sum_{i\hat{\tau}\sigma}
C^{\dagger}_{i\sigma}C_{i+\hat{\tau}\sigma}+\mu\sum_{i\sigma}
C^{\dagger}_{i\sigma}C_{i\sigma}\nonumber\\
&+&J\sum_{i\hat{\eta}}{\bf S}_{i} \cdot {\bf S}_{i+\hat{\eta}},
\end{eqnarray}
acting on the Hilbert subspace with no double occupied site, i.e.,
$\sum_{\sigma}C^{\dagger}_{i\sigma}C_{i\sigma}\leq 1$, captures the
essential physics of the doped CuO$_{2}$ plane \cite{anderson87},
where $\hat{\eta}=\pm\hat{x},\pm \hat{y}$, $\hat{\tau}=\pm\hat{x}
\pm\hat{y}$, $C^{\dagger}_{i\sigma}$ ($C_{i\sigma}$) is the electron
creation (annihilation) operator, ${\bf S}_{i}=(S^{x}_{i},S^{y}_{i},
S^{z}_{i})$ are spin operators, and $\mu$ is the chemical potential.
To deal with the constraint of no double occupancy in analytical
calculations, the charge-spin separation (CSS) fermion-spin theory
\cite{feng04} has been developed, where the constrained electron
operators are decoupled as $C_{i\uparrow}=
h^{\dagger}_{i\uparrow}S^{-}_{i}$ and $C_{i\downarrow}=
h^{\dagger}_{i\downarrow}S^{+}_{i}$, with the spinful fermion
operator $h_{i\sigma}=e^{-i\Phi_{i\sigma}}h_{i}$ describes the
charge degree of freedom together with some effects of the spin
configuration rearrangements due to the presence of the doped hole
itself (charge carrier), while the spin operator $S_{i}$ describes
the spin degree of freedom (spin), then the electron local
constraint for single occupancy, $\sum_{\sigma}C^{\dagger}_{i\sigma}
C_{i\sigma}=S^{+}_{i}h_{i\uparrow}h^{\dagger}_{i\uparrow}S^{-}_{i}
+S^{-}_{i}h_{i\downarrow}h^{\dagger}_{i\downarrow}S^{+}_{i}=h_{i}
h^{\dagger}_{i}(S^{+}_{i}S^{-}_{i}+S^{-}_{i}S^{+}_{i})=1-
h^{\dagger}_{i}h_{i}\leq 1$, is satisfied in analytical
calculations. In particular, it has been shown \cite{feng08} that
under the decoupling scheme, this CSS fermion-spin representation is
a natural representation of the constrained electron defined in the
Hilbert subspace without double electron occupancy. Furthermore,
these charge carrier and spin are gauge invariant
\cite{feng04,feng08}, and in this sense they are real and can be
interpreted as physical excitations \cite{laughlin}. This is much
different from the usual slave-boson approach \cite{palee,yu92},
where the electron operator is decomposed as the holon and spinon,
however, the local constraint for the single occupancy is explicitly
replaced by a global constraint in the mean-field level. Due to the
constraint, these holon and spinon are also coupled by a strong
gauge field \cite{palee,yu92}, allowed by this slave-boson
representation, and therefore these holon and spinon are not gauge
invariant. From this point of view, our treatment of constraint for
the physical electron may be better than the usual slave-boson
approach. In this CSS fermion-spin representation, the $t$-$J$
Hamiltonian (\ref{t-j1}) can be expressed as,
\begin{eqnarray}\label{t-j2}
H&=&t\sum_{i\hat{\eta}}(h^{\dagger}_{i+\hat{\eta}\uparrow}
h_{i\uparrow}S^{+}_{i}S^{-}_{i+\hat{\eta}}+
h^{\dagger}_{i+\hat{\eta} \downarrow}h_{i\downarrow}S^{-}_{i}
S^{+}_{i+\hat{\eta}})\nonumber\\
&-&t'\sum_{i\hat{\tau}} (h^{\dagger}_{i+\hat{\tau}\uparrow}
h_{i\uparrow}S^{+}_{i} S^{-}_{i+\hat{\tau}}+
h^{\dagger}_{i+\hat{\tau}\downarrow}h_{i\downarrow}S^{-}_{i}
S^{+}_{i+\hat{\tau}})\nonumber \\
&-&\mu\sum_{i\sigma}h^{\dagger}_{i\sigma} h_{i\sigma}+J_{{\rm eff}}
\sum_{i\hat{\eta}}{\bf S}_{i}\cdot {\bf S}_{i+\hat{\eta}},
\end{eqnarray}
with $J_{{\rm eff}}=(1-\delta)^{2}J$, and $\delta=\langle
h^{\dagger}_{i\sigma}h_{i\sigma}\rangle=\langle h^{\dagger}_{i}
h_{i}\rangle$ is the hole doping concentration. As a consequence,
the kinetic energy term in the $t$-$J$ model has been transferred as
an interaction between charge carriers and spins, which reflects
that even the kinetic energy term in the $t$-$J$ Hamiltonian has a
strong Coulombic contribution due to the restriction of no double
electron occupancy of a given site.

For the understanding of the physical properties of cuprate
superconductors in the SC state, we have developed a kinetic energy
driven SC mechanism \cite{feng03}, where the interaction between
charge carriers and spins arising directly from the kinetic energy
term in the $t$-$J$ model (\ref{t-j2}) induces a d-wave charge
carrier pairing state by exchanging spin excitations in the higher
power of the hole doping concentration, then the electron Cooper
pairs originating from the charge carrier pairing state are due to
the charge-spin recombination, and their condensation reveals the SC
ground-state. In particular, this d-wave SC state is controlled by
both the SC gap function and the quasiparticle coherent weight,
which leads to a fact that the maximal SC transition temperature
occurs around the optimal doping, and then decreases in both
underdoped and overdoped regimes. Furthermore, it has been shown
that this SC state is a conventional BCS-like with the d-wave
symmetry \cite{guo07}, so that the basic BCS formalism with a d-wave
SC gap function is still valid in quantitatively reproducing some
main low energy features of the SC coherence of quasiparticles,
although the pairing mechanism is driven by the kinetic energy by
exchanging spin excitations. Within this kinetic energy driven
superconductivity, we have discussed the low energy electronic
structure \cite{feng08,guo07} of cuprate superconductors, the
dynamical spin response \cite{feng03,cheng08}, and the quasiparticle
transport in the SC state \cite{wang09}, and qualitatively
reproduced some main features of ARPES experiments \cite{shen03},
inelastic neutron scattering experiments \cite{dai01,arai99}, and
microwave conductivity measurements \cite{harris} on cuprate
superconductors. The typical feature of this kinetic energy driven
SC mechanism is that the pairing comes out from the kinetic energy
by exchanging spin excitations and is not driven by the magnetic
superexchange interaction as in the resonant valence bond type
theories \cite{anderson87}. In particular, a possible SC theory has
been developed for description of superconductivity in doped
hexaborides \cite{balents00}, where the physical mechanism favoring
such a reorientation is the enhanced coherence (and hence lower
kinetic energy) of the doped electrons in a ferromagnetic background
relative to the paramagnet. Following the previous discussions
\cite{guo07,feng03}, the charge carrier diagonal and off-diagonal
Green's functions of the $t$-$J$ model (\ref{t-j2}) can be obtained
as,
\begin{eqnarray}
g_{11}({\bf k},\omega)&=&Z_{hF}\left ({U^{2}_{h{\bf k}}\over \omega-
E_{h{\bf k}}}+{V^{2}_{h{\bf k}} \over \omega+E_{h{\bf k}}}\right ),
\label{gg1}\\
g_{21}({\bf k},\omega)&=&-Z_{hF}{\bar{\Delta}_{hZ}({\bf k}) \over
2E_{h{\bf k}}}\left ({1\over \omega-E_{h{\bf k}}}-{1\over \omega+
E_{h{\bf k}}}\right ),~~~\label{gg2}
\end{eqnarray}
where the charge carrier quasiparticle spectrum $E_{h{\bf k}}=\sqrt
{\bar{\xi^{2}_{{\bf k}}}+\mid \bar{\Delta}_{hZ}({\bf k})\mid^{2}}$
with the renormalized d-wave charge carrier pair gap function
$\bar{\Delta}_{hZ}({\bf k})=\bar{\Delta}_{hZ}[{\rm cos}k_{x}-{\rm
cos}k_{y}]/2$, and the charge carrier quasiparticle coherence
factors $U^{2}_{h{\bf k}} =(1+\bar{\xi_{{\bf k}}}/E_{h{\bf k}})/2$
and $V^{2}_{h{\bf k}}= (1-\bar{\xi_{{\bf k}}}/E_{h{\bf k}})/2$,
while the charge carrier quasiparticle coherent weight $Z_{hF}$ and
other notations are defined as same as in Ref. \onlinecite{guo07},
and have been determined by the self-consistent calculation
\cite{guo07,feng03}. For the convenience of the following
discussions, the full charge carrier Green functions (\ref{gg1}) and
(\ref{gg2}) can also be expressed in the Nambu representation as,
\begin{eqnarray}
g({\bf{k}},\omega)=Z_{\rm{hF}}\,\frac{\omega\tau_0+
\bar{\xi}_{\bf{k}}\tau_3 - \bar{\Delta}_{\rm{hZ}}({\bf{k}})
\tau_1}{\omega^2 - E_{{\rm{h}}{\bf{k}}}^2},
\label{holegreenfunction}
\end{eqnarray}
where $\tau_{0}$ is the unit matrix, $\tau_{1}$ and $\tau_{3}$ are
Pauli matrices.

In the CSS fermion-spin representation \cite{feng04,feng08}, the
electron Green's function in the Nambu representation,
\begin{eqnarray}
G({\bf k},\omega)=\left(
\begin{array}{cccc}
G_{11}({\bf k},\omega), & G_{12}({\bf k},\omega) \\
G_{21}({\bf k},\omega), & G_{22}({\bf k},\omega)
\end{array} \right) \,, \label{electrongreenfunction}
\end{eqnarray}
is a convolution of the spin Green's function and charge carrier
Green's function (\ref{holegreenfunction}), and its diagonal and
off-diagonal components $G_{11} (i-j,t-t') =\langle\langle
C_{i\sigma}(t); C^{\dagger}_{j\sigma} (t')\rangle \rangle$ and
$G_{21}(i-j,t-t')=\langle \langle C^{\dagger}_{i\uparrow}(t);
C^{\dagger}_{j\downarrow}(t')\rangle \rangle$ have been obtained as
\cite{guo07},
\begin{widetext}
\begin{subequations}
\begin{eqnarray}
G_{11}({\bf k},\omega)&=&{1\over N}\sum_{{\bf p}}Z_{F}{B_{{\bf p}}
\over 2\omega_{{\bf p}}}\left \{ {\rm coth}[{1\over 2}\beta
\omega_{{\bf p}}]\left ({U^{2}_{h{\bf p+k}}\over\omega+E_{h{\bf
p+k}}-\omega_{{\bf p}}}+{U^{2}_{h{\bf p+k}}\over \omega+E_{h{\bf
p+k}}+\omega_{{\bf p}}}\right . \right .\nonumber \\
&+&\left . {V^{2}_{h{\bf p+k}}\over \omega-E_{h{\bf p+k}}+
\omega_{{\bf p}}}+{V^{2}_{h{\bf p+k}}\over \omega-E_{h{\bf p+k}}
-\omega_{{\bf p}}} \right )+{\rm tanh}[{1\over 2}\beta E_{h{\bf
p+k}}]\left ({U^{2}_{h{\bf p+k}}\over \omega +E_{h{\bf p+k}}
+\omega_{{\bf p}}}\right .\nonumber\\
&-&\left .\left . {U^{2}_{h{\bf p+k}}\over \omega+E_{h{\bf
p+k}}-\omega_{{\bf p} }}+{V^{2}_{h{\bf p+k}}\over \omega-E_{h{\bf
p+k}}- \omega_{{\bf p}}}-{V^{2}_{h{\bf p+k}}\over \omega-E_{h{\bf
p+k}}+\omega_{{\bf p}}} \right ) \right \} ,\\
G_{21}({\bf k},\omega)&=&{1\over N}\sum_{{\bf p}} Z_{F}
{\bar{\Delta}_{hZ}({\bf p+k})\over 2E_{h{\bf p+k}}}{B_{{\bf p }}
\over 2\omega_{{\bf p}}}\left \{{\rm coth}[{1\over 2}\beta
\omega_{{\bf p}}]\left ({1\over \omega-E_{h{\bf p+k}} -\omega_{{\bf
p}}}+{1\over \omega-E_{h{\bf p+k}}+\omega_{{\bf p}}}\right .\right .
\nonumber \\
&-&\left . {1\over \omega +E_{h{\bf p+k}}+\omega_{{\bf p}}}- {1\over
\omega +E_{h{\bf p+k}}-\omega_{{\bf p}}} \right )+{\rm tanh}[{1\over
2}\beta E_{h{\bf p+k}}]\left({1\over \omega
-E_{h{\bf p+k}}-\omega_{{\bf p}}}\right .\nonumber\\
&-&\left .\left .{1\over \omega-E_{h{\bf p+k}} +\omega_{{\bf p}}} -
{1\over \omega +E_{h{\bf p+k}}+\omega_{{\bf p} }}+ {1\over \omega
+E_{h{\bf p+k}}-\omega_{{\bf p}}} \right )\right \},
\end{eqnarray}
\end{subequations}
\end{widetext}
respectively, where the electron quasiparticle coherent weight
$Z_{F}=Z_{hF}/2$, and the spin excitation spectrum $\omega_{{\bf
p}}$ and $B_{\bf p}$ have been given in Ref. \onlinecite{guo07}.

In the presence of a single point-like impurity scattering
potential,
\begin{eqnarray}
\tilde{V}=V_0\delta({\bf r})\tau_3,
\label{potential}
\end{eqnarray}
the unperturbed electron Green's function in Eq.
(\ref{electrongreenfunction}) is dressed by this impurity
scattering, where the $T$ matrix exactly accounts for multiple
scattering off that impurity \cite{balatsky06}. Since translational
invariance is broken by the impurity, the dressed electron Green's
function in the Nambu representation in real-space depends on two
positions ${\bf r}$ and ${\bf r}'$,
\begin{eqnarray}
\tilde{G}({\bf r},{\bf r}',\omega)=G({\bf r}-{\bf r}',\omega)+
G({\bf r},\omega)\tilde{T}(\omega)G(-{\bf r}',\omega),
\label{dressedgreenfunction}
\end{eqnarray}
with the impurity induced $T$ matrix can be obtained as
\cite{balatsky06},
\begin{eqnarray}
\tilde{T}(\omega)= {1\over 1-G(\omega)V_0\tau_3}V_0{\tau_3},
\label{T-matrix}
\end{eqnarray}
where $G(\omega)=(1/N)\sum_{{\bf k}}G({\bf k},\omega)$. In this
case, LDOS of the system can be obtained as,
\begin{eqnarray}
\rho({\bf r},\omega)=-{1\over\pi}{\rm Im}\tilde{G}({\bf r},\omega)
=\rho_{0}(\omega)+\delta\rho({\bf r},\omega),
\label{density1}
\end{eqnarray}
where the homogeneous density of states $\rho_{0}(\omega)=- {\rm Im}
G(0,\omega)/\pi$, and is uniform in real-space, therefore it
reflects a homogenous background, while $\delta\rho({\bf r},\omega)$
is the modulation for the homogeneous density of states due to the
presence of the impurity scattering potential (\ref{potential}), and
can be obtained as,
\begin{eqnarray}
\delta\rho({\bf r},\omega)&=&\rho({\bf r},\omega)-\rho_{0}(\omega)
\nonumber\\
&=& -{1\over\pi}{\rm Im}[G({\bf r},\omega)\tilde{T}(\omega)G(-{\bf
r}, \omega)]_{11}, \label{density2}
\end{eqnarray}
and its Fourier transform is evaluated explicitly as,
\begin{eqnarray}
\delta\rho({\bf q},\omega)=-{1\over\pi}{\rm Im}\left ( {1\over N}
\sum_{{\bf k}}[G({\bf k}+{\bf q},\omega)\tilde{T}(\omega)G({\bf k},
\omega)]_{11}\right ) .~~ \label{density3}
\end{eqnarray}
This LDOS is closely related to the differential conductance $g({\bf
r},\omega)$ since the result of the differential conductance $g({\bf
r},\omega)$ is proportional to $\rho({\bf r},\omega)$ at location
${\bf r}$ and energy $\omega$=eV. However, the intense atomic-scale
spatial fluctuations in electronic structure cause systematic errors
in setting the STM tip elevation for the experimental measurement of
the differential conductance $g({\bf r},\omega)$
\cite{kohsaka07,hanaguri07,kohsaka08}, therefore the ratio of
differential conductances at opposite bias in Eq. (\ref{ratio}) or
its equivalent, the ratio of LDOS at opposite energy,
\begin{eqnarray}
Z({\bf r},\omega)={\rho({\bf r},\omega)\over\rho({\bf r},-\omega)},
\label{ratio1}
\end{eqnarray}
has been measured experimentally for an enhancement of the QSI
signatures  \cite{kohsaka07,hanaguri07,kohsaka08}. Since the
condition $\rho_0(\omega)\gg\delta\rho({\bf r} , \omega)$ is well
satisfied for cuprate superconductors, then the ratio of LDOS at
opposite energy (\ref{ratio1}) can be obtained approximately as
\cite{franz08},
\begin{eqnarray}
Z({\bf r},\omega)\approx Z_0(\omega)\left [1+{\delta\rho({\bf r},
\omega) \over\rho_0(\omega)}-{\delta\rho({\bf r},-\omega)\over
\rho_0(-\omega)}\right ],
\end{eqnarray}
where $Z_0(\omega)=\rho_0(\omega)/\rho_0(-\omega)$, and only the
first-order modulation $\delta\rho({\bf r},\pm\omega)$ is kept, then
the modulation of the ratio of LDOS at opposite energy can be
expressed as,
\begin{eqnarray}
\delta Z({\bf r},\omega)&=&Z({\bf
r},\omega)-Z_0(\omega)\nonumber\\
&\approx& Z_{0}(\omega) \left [{\delta\rho ({\bf r},\omega)\over
\rho_0(\omega)}-{\delta\rho({\bf r},-\omega) \over\rho_0(-\omega)}
\right ],
\end{eqnarray}
and its Fourier transformation is evaluated explicitly as,
\begin{eqnarray}\label{equ-z}
\delta Z({\bf q},\omega)\approx Z_{0}(\omega)\left [{\delta\rho
({\bf q},\omega)\over\rho_0(\omega)}-{\delta\rho({\bf q},-\omega)
\over\rho_0(-\omega)}\right ].
\end{eqnarray}
It has been argued \cite{kohsaka07,hanaguri07,kohsaka08} that the
observed QSI from $\delta Z({\bf r}, \omega)$ or $\delta Z({\bf
q},\omega)$ is an intrinsic phenomenon free from any
set-point-related issues, which inevitably contaminate $g({\bf
r},\omega)$ and $g({\bf q},\omega)$.

\begin{figure}[h!]
\includegraphics[scale=0.35]{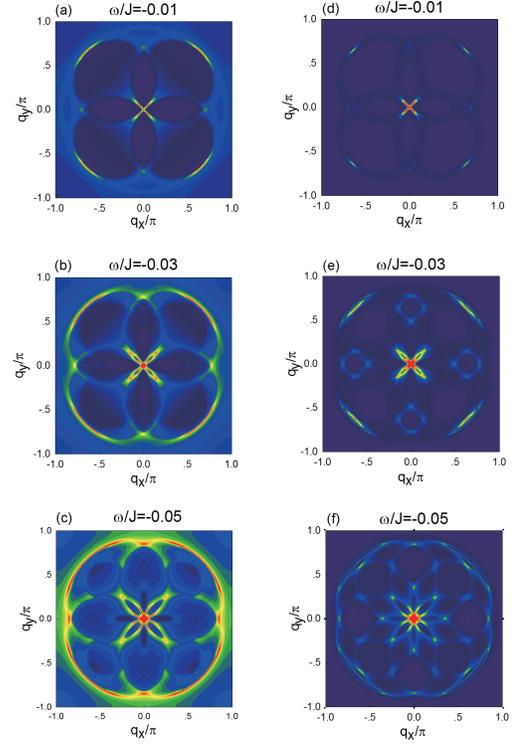}
\caption{(Color online) The Fourier transformed LDOS as a function
of momentum in the full Brillouin zone with energy (a)
$\omega=-0.01J$, (b) $\omega=-0.03J$, and (c) $\omega=-0.05J$ at
temperature $T=0.002J$ for the doping concentration $\delta=0.15$ in
the presence of single point-like potential scatterer of strength
$V=0.1J$ (left panel). (d) to (f) are the corresponding results of
the Fourier transformed ratio of LDOS at opposite energy (right
panel).}
\end{figure}

We are now ready to discuss the QSI phenomenon in cuprate
superconductors in the SC state and its extinction at the high
energy. In cuprate superconductors, although the values of $J$ and
$t$ is believed to vary somewhat from compound to compound, however,
as a qualitative discussion, the commonly used parameters in this
paper are chosen as $t/J=2.5$ and $t'/t=0.3$. In this case, we have
performed a calculation for the Fourier transformed LDOS
$\delta\rho({\bf q}, \omega)$ in Eq. (\ref {density3}) and the
Fourier transformed ratio of LDOS at opposite energy $\delta Z({\bf
q},\omega)$ in Eq. (\ref{equ-z}), and the results of
$|\delta\rho({\bf q},\omega)|$ as a function of momentum in the full
Brillouin zone with energy (a) $\omega=-0.01J$, (b) $\omega=-0.03J$,
and (c) $\omega=-0.05J$ at temperature $T=0.002J$ for the doping
concentration $\delta=0.15$ in the presence of single point-like
potential scatterer of strength $V=0.1J$ are plotted in Fig. 1 (left
panel). For comparison, the corresponding results of $|\delta Z({\bf
q},\omega)|$ are also plotted in Fig. 1d-f (right panel). It is
shown that the results obtained from $\delta Z({\bf q},\omega)$ are
almost the same as those of $\delta\rho({\bf q},\omega)$, reflecting
an experimental fact that the measurement data from the ratio of
LDOS at opposite energy retain all the main information of QSI
observed from LDOS. Moreover, both results from $\delta Z({\bf
q},\omega)$ and $\delta\rho({\bf q}, \omega)$ are clearly fourfold
symmetric and display numerous local maxima (bright regions) at
different ${\bf q}$ for different energies, where all of these ${\bf
q}$ vectors are consistent with the prediction from the octet model
\cite{hoffman02,mcelroy03}. These bright regions in momentum-space
display the intensity of the modulation for LDOS (Fig. 1a-c) or the
intensity of the modulation for the ratio of LDOS at opposite energy
(Fig. 1d-f). In particular, the bright regions near the center of
each figure in Fig. 1 a-c or Fig. 1d-f reflect long-wavelength
inhomogeneity in $\delta\rho({\bf q},\omega)$ or $\delta Z({\bf
q},\omega)$, and are induced by the weak impurity scattering
potential (\ref{potential}). All these results are in qualitative
agreement with the STM experimental results
\cite{hoffman02,mcelroy03}.

\begin{figure}[h!]
\includegraphics[scale=0.45]{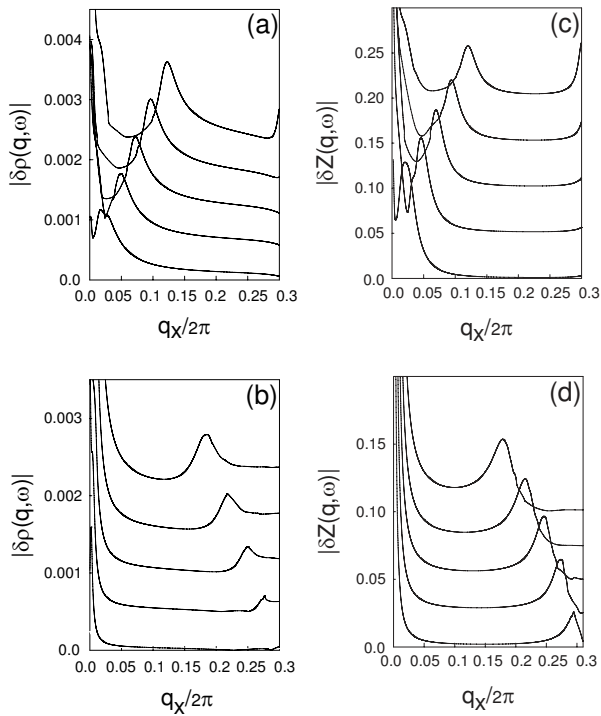}
\caption{The Fourier transformed LDOS as a function of momentum
along the (a) $[0,0]\rightarrow [\pi,\pi]$ direction and (b)
$[0,0]\rightarrow [\pi,0]$ direction with energy $\omega= -0.01J$,
$\omega=-0.02J$, $\omega=-0.03J$, $\omega=-0.04J$, and
$\omega=-0.05J$ (from bottom to top) at temperature $T=0.002J$ for
the doping concentration $\delta=0.15$ in the presence of single
point-like potential scatterer of strength $V=0.1J$ (left panel).
(c) and (d) are the corresponding results of the Fourier transformed
ratio of LDOS at opposite energy (right panel).}
\end{figure}

\begin{figure}[h!]
\includegraphics[scale=0.45]{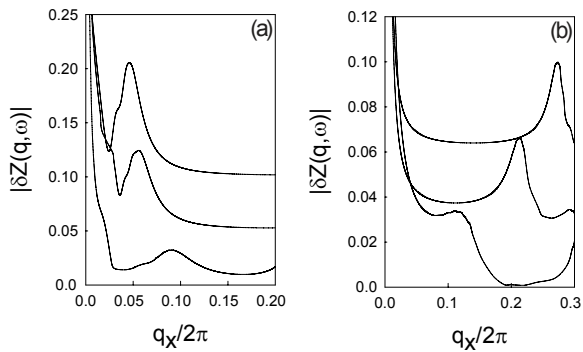}
\caption{The Fourier transformed ratio of LDOS at opposite energy as
a function of momentum along the (a) $[0,0]\rightarrow [\pi,\pi]$
direction and (b) $[0,0]\rightarrow [\pi,0]$ direction with energy
$\omega= -0.02J$ at temperature $T=0.002J$ for the doping
concentration $\delta=0.08$, $\delta=0.11$, and $\delta=0.15$ (from
bottom to top) in the presence of single point-like potential
scatterer of strength $V=0.1J$.}
\end{figure}

To analyze this peak feature in Fig. 1 more clearly, we have made a
series of calculations for the momentum dependence of
$\delta\rho({\bf q},\omega)$ and $\delta Z({\bf q},\omega)$ with
different energies, and the results of $|\delta\rho({\bf
q},\omega)|$ as a function of momentum along the (a)
$[0,0]\rightarrow [\pi,\pi]$ (nodal) direction and (b)
$[0,0]\rightarrow [\pi,0]$ (antinodal) direction with energy
$\omega= -0.01J$, $\omega=-0.02J$, $\omega=-0.03J$, $\omega=-0.04J$,
and $\omega=-0.05J$ (from bottom to top) at temperature $T=0.002J$
for the doping concentration $\delta=0.15$ in the presence of single
point-like potential scatterer of strength $V=0.1J$ are plotted in
Fig. 2 (left panel). For comparison, the corresponding results of
$|\delta Z({\bf q},\omega)|$ are also plotted in Fig. 2c-d (right
panel). In corresponding to the results in Fig. 1, the results in
Fig. 2 show that although the peak in $\delta\rho({\bf q},\omega)$
(then the intensity of the modulation for LDOS) or the peak in
$\delta Z({\bf q}, \omega)$ (then the intensity of the modulation
for the ratio of LDOS at opposite energy) is changing rapidly and
intricately with energy, the peaks, whose ${\bf q}$-vectors oriented
towards the antinodal direction, appear at finite $|{\bf q}|$ at
very low energy and then move steadily inwards toward the center,
while the peaks with ${\bf q}$-vectors along the nodal direction,
appear and move steadily to large $|{\bf q}|$ with increasing
energy, in qualitative agreement with the STM experimental results
\cite{hoffman02,mcelroy03}. However, although the position of the
peak obtained from $\delta\rho({\bf q}, \omega)$ is almost the same
as that from $\delta Z({\bf q},\omega)$, the peak intensity obtained
from $\delta Z({\bf q},\omega)$ is much stronger than that appeared
in $\delta\rho({\bf q},\omega)$, and therefore there is an
enhancement of the QSI signatures in $\delta Z({\bf q},\omega)$.
This expected difference of the peak intensity obtained from
$\delta\rho({\bf q},\omega)$ and $\delta Z({\bf q},\omega)$ can be
understood from the definition of $\delta Z({\bf q},\omega)$ in Eq.
(\ref{equ-z}). Although the modulations in $\delta\rho({\bf r},
\omega)$ and $\delta\rho({\bf r},-\omega)$ are spatially quite
similar, the spatial-phase relation for QSI is not known precisely.
However, $\delta Z({\bf q},\omega)$ is sensitive to the relation
signs between $\delta\rho({\bf q},\omega)$ and $\delta\rho({\bf q},
-\omega)$, namely, taking the ratio of LDOS reduces the in-plane
component \cite{hanaguri07}. This is why the the QSI extinction at
the high energy in cuprate superconductors in the SC state is
firstly observed from the experimental measurement of $\delta Z({\bf
q},\omega)$ \cite{kohsaka07,hanaguri07,kohsaka08}.

For a better understanding of the physical properties of the QSI
phenomenon in cuprate superconductors, we have further performed a
calculation for $\delta Z({\bf q},\omega)$ with different doping
concentrations, and the results of $|\delta Z({\bf q},\omega)|$ as a
function of momentum along the (a) $[0,0]\rightarrow [\pi,\pi]$
(nodal) direction and (b) $[0,0]\rightarrow [\pi,0]$ (antinodal)
direction with energy $\omega= -0.02J$ at temperature $T=0.002J$ for
the doping concentration $\delta=0.08$, $\delta=0.11$, and
$\delta=0.15$ (from bottom to top) in the presence of single
point-like potential scatterer of strength $V=0.1J$ are plotted in
Fig. 3. Obviously, the peak is doping dependent. With increasing the
doping concentration, the weight of the peak increases. Furthermore,
the position of the peak along the nodal direction appears at finite
$|{\bf q}|$ in the low doping concentration, and then moves towards
to the center of the Brillouin zone with increasing doping
concentration. In contrast to the case along the nodal direction,
the position of the peak along the antinodal direction is located at
finite $|{\bf q}|$ in the low doping concentration, and then is
shifted to large $|{\bf q}|$ with increasing doping concentration.

\begin{figure}[h!]
\includegraphics[scale=0.35]{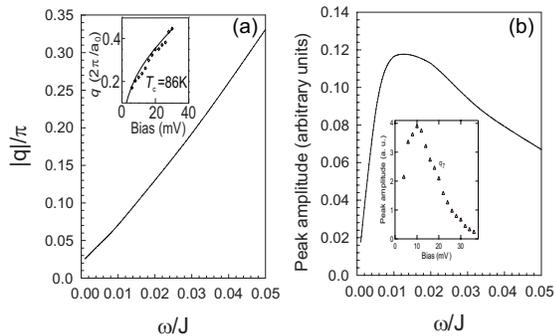}
\caption{(a) The positions and (b) amplitudes of the lowest energy
peaks in the Fourier transformed ratio of LDOS at opposite energy as
a function of energy along the $[0,0]\rightarrow [\pi,\pi]$
direction for temperature $T=0.002J$ at the doping concentration
$\delta=0.15$ in the presence of single point-like potential
scatterer of strength $V=0.1J$. Insets: the corresponding
experimental results for cuprate superconductors taken from Ref.
\onlinecite{kohsaka08}.}
\end{figure}

\begin{figure}[h!]
\includegraphics[scale=0.35]{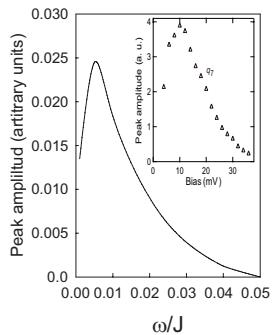}
\caption{The amplitudes of the lowest energy peaks in the Fourier
transformed ratio of LDOS at opposite energy as a function of energy
along the $[0,0]\rightarrow [\pi,\pi]$ direction with temperature
$T=0.002J$ at the doping concentration $\delta=0.15$ for parameters
$\Gamma=0.005J$ and $\alpha=0.2$ in the presence of single
point-like potential scatterer of strength $V=0.1J$. Inset: the
corresponding experimental result for cuprate superconductors taken
from Ref. \onlinecite{kohsaka08}. }
\end{figure}

The energy dependence of the peaks in Fig. 2 has been used to
extract the dispersion of the peaks, and the results of (a) the
positions and (b) amplitudes of the lowest energy peaks in $|\delta
Z({\bf q},\omega)|$ as a function of energy along the
$[0,0]\rightarrow [\pi,\pi]$ direction for temperature $T=0.002J$ at
the doping concentration $\delta=0.15$ in the presence of single
point-like potential scatterer of strength $V=0.1J$ are plotted in
Fig. 4. For comparison, the corresponding STM experimental results
\cite{kohsaka08} of cuprate superconductors are also shown in Fig. 4
(inset). Our results show clearly that the sharp peak persists in a
large momentum space region along the nodal direction and are in
qualitative agreement with the STM experimental result
\cite{kohsaka08}. However, the amplitude of the peak increases with
increasing energy at the low energy, and reaches a maximum at the
intermediate energy, then decreases rapidly with further increasing
energy. Apparently, there is a substantial difference between theory
and experiment, namely, the amplitude of the peak obtained from
theory does not diminish to zero at the high energy. This obvious
weakness is a natural consequence of the neglect of the imaginary
part of the self-energy (then the scattering rate). To obtain the
full charge carrier Green function (\ref{holegreenfunction}) within
the kinetic energy driven SC mechanism, we \cite{feng03,guo07} have
made a self-consistent calculation in the static limit approximation
for the real part of the charge carrier self-energy ${\rm Re}
\Sigma^{(h)}_{1}(\omega)$ induced by the interaction between the
charge carriers and spins, where its antisymmetric part has been
obtained as $Z^{-1}_{hF}=1- {\rm Re} \Sigma^{(h)}_{1o}({\bf k},
\omega=0)\mid_{{\bf k}=[\pi,0]}$ and therefore is closely related to
the charge carrier quasiparticle coherent weight, while its
symmetric part ${\rm Re}\Sigma^{(h)}_{1e}({\bf k},\omega=0)$ is a
constant around the Fermi surface, and it just renormalizes the
chemical potential. However, the imaginary part of the charge
carrier self-energy ${\rm Im}\Sigma^{(h)}_{1}({\bf k},\omega)$ has
been dropped \cite{feng03,guo07}. This leads to a fact that the
imaginary part of the renormalized electron self-energy in the
dressed electron Green's function (\ref{dressedgreenfunction}) due
to the presence of the impurity scattering potential
(\ref{potential}) has been neglected. In this case, for the present
discussions of the QSI extinction at the high energy, we need to
introduce an imaginary part of the electron self-energy (then the
scattering rate) in the electron Green function as,
\begin{eqnarray}
i{\rm Im}\Sigma(\omega)=i(\Gamma+\alpha\omega),
\label{scattering-rate}
\end{eqnarray}
for a compensation due to the neglect of the imaginary part of the
renormalized self-energy, where $\Gamma$ and $\alpha$ are constant.
It has been argued \cite{alldredge08} that this type of the energy
dependent inelastic scattering rate (\ref{scattering-rate}) seems to
be an intrinsic property of the electronic structure of cuprate
superconductors, since it has been used to provide a robust fit for
the spatially inhomogeneous differential conductances of cuprate
superconductor based on the phenomenological d-wave BCS formalism.
By considering this scattering rate (\ref{scattering-rate}) in the
electron Green function, we therefore find that the dispersing QSI
always disappears at the high energy. To show this point clearly, we
have made a series of calculations for $\delta Z({\bf q},\omega)$
with different momenta and energies, and the result of the
amplitudes of the lowest energy peaks in $|\delta Z({\bf q},
\omega)|$ as a function of energy along the $[0,0]\rightarrow
[\pi,\pi]$ direction for temperature $T=0.002J$ at the doping
concentration $\delta=0.15$ for parameters $\Gamma=0.005J$ and
$\alpha=0.2$ in the presence of single point-like potential
scatterer of strength $V=0.1J$ is plotted in Fig. 5 in comparison
with the corresponding STM experimental result \cite{kohsaka08} of
cuprate superconductors (inset). In comparison with the result in
Fig. 4b, although the amplitude of the peak is severely suppressed
due to the presence of the scattering rate (\ref{scattering-rate}),
it diminishes to zero (then the QSI extinction) at the high energy.
Moreover, during the calculations, we find that the constant
scattering rate $\Gamma$ in Eq. (\ref{scattering-rate}) only plays a
subsidiary role, while the QSI modulation is further weakened by the
effective scattering rate $\alpha\omega$. The similar conclusion has
been obtained in Ref. \onlinecite{alldredge08} where they also find
that the effective scattering rate $\alpha\omega$ plays a key role.
Using a reasonably estimative value of $J\approx 120\sim 150$ meV,
the peak disappears around the energy 6 meV$\sim$7.5 meV.
Apparently, there is still a substantial difference between theory
and experiment, namely, the energy value of the QSI extinction
calculated theoretically is smaller than the corresponding energy
value measured in the experiment. However, upon a closer examination
one can see immediately that the main difference is due to fact that
the calculated peak energy decreases rapidly with energy at high
energy. This energy value of the QSI extinction is also smaller than
the corresponding value of the charge carrier pairing gap
\cite{guo07}. The simple $t$-$J$ model can not be regarded as a
complete model for the quantitative comparison with cuprate
superconductors, however, as for a qualitative discussion in this
paper, the overall shape seen in the theoretical result is
qualitatively consistent with that observed in the STM experiment
\cite{kohsaka08}.

A nature question is why QSI below the high energy in cuprate
superconductors and its extinction at the high energy can be
described qualitatively in the framework of the kinetic energy
driven SC mechanism in the presence of a single point-like impurity.
This may be understood from the kinetic energy driven SC mechanism
itself \cite{feng03}. As we have mentioned above, in the framework
of kinetic energy driven SC mechanism, the electron Cooper pairs
originating from the d-wave charge carrier pairing state are due to
the charge-spin recombination, therefore there is a coexistence of
the electron Cooper pair and short-range AF correlation, and hence
the short-range AF fluctuation can persist into the SC state
\cite{feng03}. In particular, this charge-spin recombination is
characterized by a convolution of the spin Green's function and
charge carrier Green's function \cite{feng03}. The main consequence
of this convolution is that the hole-like charge carrier d-wave BCS
formalism is transferred into the electron-like d-wave BCS formalism
\cite{guo07}. In other words, main difference between the hole-like
charge carrier Green's function (\ref{holegreenfunction}) and the
electron-like Green's function (\ref{electrongreenfunction}) is a
shift of the momentum by the AF wave vector $Q=[\pi,\pi]$ in the
Brillouin zone \cite{guo07}. This leads to a fact that the
quasiparticle states near the antinodal points are broadened.
Moreover, these quasiparticle states around the antinodal regime are
further suppressed due to the presence of the impurity scattering
potential ({\ref{potential}). On the other hand, the kinetic energy
driven d-wave SC state is controlled by both the SC gap function and
the SC quasiparticle coherent weight, which indicates that only
coherent Bogoliubov quasiparticles at the Fermi surface are
available for superconductivity since everything happens at the
Fermi surface. In this case, although the Bogoliubov quasiparticles
can be excited by either positive or negative energies
\cite{wang10}, the excitation spectrum is a particle-hole symmetric
thin Dirac cone around the nodal regime, then QSI below the high
energy and its extinction at the high energy occur in this case,
where this high energy which sets the QSI extinction is determined
not only by the d-wave SC gap function alone but also by the
coherent Bogoliubov quasiparticles located on the thin Dirac cone.

In summary, within the framework of the kinetic energy driven SC
mechanism, we have discussed the QSI phenomenon in cuprate
superconductors in the SC state by considering the single point-like
impurity scattering potential. This QSI is characterized by the
peaks in LDOS or the ratio of LDOS at opposite energy. By
calculation of the momentum and energy dependence of the Fourier
transformed LDOS $\delta\rho({\bf q},\omega)$ and the Fourier
transformed ratio of LDOS at opposite energy $\delta Z({\bf
q},\omega)$, we have shown that the remarkable QSI phenomenon
observed from STM experiments on cuprate superconductors can be
described qualitatively by a single point-like impurity in the
kinetic energy driven homogeneous d-wave SC state. The amplitude of
the peak increases with increasing energy at the low energy, and
reaches a maximum at the intermediate energy, then diminishes to
zero at the high energy. The theory also predicts that with
increasing doping concentration, the position of the peak along the
nodal direction moves towards to the center of the Brillouin zone,
while the position of the peak along the antinodal direction is
shifted to large momentum region, which should be verified by
further STM experiments.

Within the framework of the kinetic energy driven SC mechanism
\cite{feng03}, our present results of QSI below the high energy and
its extinction at the high energy due to the presence of the
impurity scattering potential (\ref{potential}) are qualitatively
consistent with the recent STM experimental data
\cite{hoffman02,mcelroy03,kohsaka08}. Establishing this agreement is
important to confirming the nature of a single impurity in the
kinetic energy driven homogeneous d-wave SC state in cuprate
superconductors. Although a quantitative understanding of the QSI
phenomenon is not straightforward to obtain, and depends rather
sensitively on the nature of the scattering medium, in this paper we
are primarily interested in exploring the general notion of the QSI
phenomenon induced by a single impurity in the kinetic energy driven
homogeneous d-wave SC state. The qualitative agreement between the
present theoretical results and STM experimental data also show that
the presence of impurities plays a crucial role for the QSI
phenomenon in cuprate superconductors in the SC state. Although this
QSI extinction at the high energy \cite{kohsaka08} also can be
fitted by using a phenomenological d-wave BCS formalism with the
same imaginary part of the self-energy (\ref{scattering-rate}),
however, no explicit calculations of the doping dependence of
$|\delta Z({\bf q},\omega)|$, as shown in Fig. 3, can be made within
a phenomenological d-wave BCS formalism.

Under the kinetic energy driven SC mechanism, the external magnetic
field aligns the spins of the unpaired electrons, then the singlet
charge carrier pairs can not take advantage of the lower energy
offered by a spin-polarized state. In this case, the magnetic field
dependence of the penetration depth and superfluid density have been
studied recently \cite{huang10}, where the superfluid density
decreases with increasing magnetic field, in agreement with the
experimental results \cite{sonier99}. With this study
\cite{huang10}, we therefore can predict that QSI should be magnetic
field dependent, and then the energy value of the QSI extinction
decreases with increasing magnetic field. One of the typical
features of d-wave superconductivity in cuprate superconductors is
the particle-hole symmetric octet of dispersive Bogoliubov
quasiparticle interference modulations
\cite{pan01,hoffman02,mcelroy03}. However, the recent STM
experiments \cite{lee09} have reported on this octet's evolution
from low temperatures to well into the underdoped pseudogap regime,
where no pronounced changes occur in the octet phenomenology at the
superconductor's critical temperature $T_{c}$, and it survives up to
at least temperature $T\sim 1.5T_{c}$. Thus an important issue is
how this octet's evolution from the SC state into the pseudogap
regime is fitted within the framework of the kinetic energy driven
superconductivity. These and the related issues are under
investigation now.

\acknowledgments

The authors would like to thank H. S. Zhao for the helpful
discussions. This work was supported by the National Natural Science
Foundation of China under Grant No. 10774015, and the funds from the
Ministry of Science and Technology of China under Grant Nos.
2006CB601002 and 2006CB921300.

\end{document}